# Determinantes concorrenciais dos atrasos dos voos no aeroporto e na rota

William Eduardo Bendinelli
Alessandro V. M. Oliveira⇀
Instituto Tecnológico de Aeronáutica, São José dos Campos, Brasil
⇀ Autor correspondente. Instituto Tecnológico de Aeronáutica. Praça Marechal Eduardo Gomes, 50. 12.280-250 - São José dos Campos, SP - Brasil.
E-mail: alessandro@ita.br.

**Resumo**: Atrasos de voos são uma realidade na indústria aérea moderna no mundo todo. Entretanto, os estudos da literatura têm investigado os determinantes concorrenciais dos atrasos advindos de fatores originários no aeroporto e na rota de forma separada. Este trabalho tem como objetivo apresentar um estudo nacional que utilizou uma abordagem unificadora da literatura, considerando os efeitos locais e globais da concorrência sobre os atrasos. A análise levou em consideração um fenômeno conhecido como "internalização das externalidades" do congestionamento no aeroporto. Além disso, discute-se a relação entre qualidade e concorrência na rota e os impactos da entrada de uma empresa aérea de baixo custo (LCC) nos atrasos das empresas aéreas incumbentes no mercado aéreo brasileiro. A literatura sugere que há evidências de internalização do congestionamento nos aeroportos brasileiros, em paralelo com a concorrência por qualidade ao nível da rota. Adicionalmente, a concorrência potencial causada pela presença da LCC provoca um efeito global que sugere a existência de impactos outros que não em preços em rotas onde ela não entrou.

*Palavras-chave*: transporte aéreo, companhias aéreas, atrasos de voo.

## I. Introdução

Atrasos de empresas aéreas tornaram-se uma realidade constante na indústria de transporte aéreo moderna em todo o mundo, de tal forma que os passageiros estão cada vez mais familiarizados com atrasos pelo menos de magnitude moderada em sua rotina de viagens. Além disso, não é incomum encontrarmos episódios de perturbações relativamente intensas nos cronogramas das companhias aéreas devido às condições meteorológicas adversas, greves ou congestionamentos, em vários lugares do mundo como, por exemplo, nos Estados Unidos, Europa e China. Por serem estressantes aos passageiros e envolverem altos custos operacionais relacionados ao congestionamento, mas também à sua mitigação, as motivações que as empresas aéreas têm para manterem a pontualidade dos voos dependem, assim, de incentivos econômicos que são muitas vezes conflitantes.

Em 2013, no Reino Unido, uma taxa de congestionamento estava sendo considerada para reduzir a demanda em dois dos mais importantes aeroportos de Londres - Heathrow e Gatwick -, e encorajar os passageiros a voarem a partir de outros aeroportos metropolitanos, como Luton e Stansted . Tal iniciativa revela como as autoridades e os operadores aeroportuários gerenciam os congestionamentos e suas consequentes perturbações aéreas sob um cenário de não-expansão do aeroporto em um futuro próximo. No Brasil, desde o final dos anos 1990, as autoridades vêm adotando a regulação do Aeroporto de São Paulo/Congonhas para conter a pressão gerada pelo rápido crescimento da demanda e os gargalos de infraestrutura apresentados desde o final dos anos 2000 até um período recente, pré-pandemia da Covid-19.

A liberalização econômica do setor aéreo promoveu mudanças estruturais significativas na indústria como um todo, através de privatizações, alianças estratégicas entre empresas aéreas, concentração da malha e de empresas, reorganização dos serviços com a introdução de inovações tecnológicas e estruturais. Como resultado destas medidas de liberalização econômica, o transporte aéreo brasileiro tem atraído novos segmentos de consumidores e se popularizado. Entre 2002 e 2010, de acordo com o Anuário Estatístico da Agência Nacional de Aviação Civil (ANAC), o transporte aéreo doméstico apresentou um aumento na demanda de 153% e uma queda nos preços de 52%.

Outro resultado do processo de liberalização econômica do transporte aéreo foi o surgimento de redes "hub-and-spoke". Pelo mundo afora, houve a formação e consolidação de aeroportos "hubs" – isto é, concentradores de tráfego para conexões. No processo de constituição de aeroportos hub, muitas empresas aéreas ganharam posição de elevada dominância sobre aeroportos importantes – como os aeroportos norte-americanos de Atlanta/Hartsfield-Jackson, pela Delta Airlines, e de Dallas/Fort Worth, pela American Airlines, dentro de suas estratégias de configuração de malha aérea. No Brasil, o Aeroporto de Brasília (BSB), por exemplo, é reconhecido como um aeroporto "hub" devido à sua localização geográfica. Igualmente, é operado na forma de hub o Aeroporto de São Paulo/Guarulhos (GRU), por estar próximo a um imenso polo gerador de viagens domésticas e internacionais (Região Metropolitana de São Paulo) e por ser um portão de entrada internacional ("gateway") natural, além da sua boa posição geográfica no que diz respeito aos países do Cone-Sul. No país, o fenômeno da dominância de aeroportos por uma empresa aérea "hub-and-spoke" é realidade mais recente e ainda longe de se assemelhar com a experiência norte-americana. Tal fato é resultado do processo de controle histórico estatal dos aeroportos e da presença de poucos aeroportos principais no país. Complementarmente, a maioria das empresas aéreas desejam em geral operar nos mesmos aeroportos, não havendo uma dispersão geográfica tão elevada. Entretanto, não se pode deixar de enfatizar a tendência à concentração de mercado e à duopolização do transporte aéreo no país, o que, com a recente crise da Covid-19, pode acirrar a questão da concorrência nos aeroportos.

## II. Revisão da Literatura

Constata-se que os estudos científicos voltados ao setor de transporte aéreo têm investigado os determinantes concorrenciais dos atrasos no aeroporto e na rota de forma separada. No primeiro caso, seguindo Daniel (1995), a literatura tem inspecionado os efeitos globais da concentração e da dominância sobre os atrasos de voos. Investigando a hipótese de internalização do congestionamento nos aeroportos, a empresa aérea dominante poderia naturalmente internalizar os custos do congestionamento que suas aeronaves impõem sobre si mesma sem a necessidade de uma tarifa de congestionamento para o caso de monopólio. Nesse sentido, a empresa aérea dominante estaria levando em consideração o custo marginal social de sua decisão. Os principais representantes dessa vertente são: Brueckner (2002), Mayer e Sinai (2003), Daniel e Harback (2008), Rupp (2009), Zhang e Zhang (2006) e Ater (2012).

Em adição, em 2014, o relatório da Agência Federal de Aviação (FAA) mostrou que a falta de concorrência nas rotas seria uma das causas do aumento das taxas de atrasos e cancelamentos dos voos das empresas aéreas . Nesse sentido, o relatório sugere que a concorrência e a qualidade do serviço prestado em termos de atrasos seriam positivamente correlacionadas e, portanto, mais frequente e maior seria a duração dos atrasos de voos em rotas menos competitivas. Estes resultados estão de acordo outra vertente da literatura de atrasos que investiga a relação entre concorrência e qualidade do serviço prestado. Iniciada por Suzuki (2000), essa literatura apresenta vários artigos econométricos recentes investigando os determinantes de atrasos ao nível da rota, tais como Mazzeo (2003), Rupp, Owens e Plumly (2006) e Greenfield (2014). Mais recentemente, alguns artigos dessa literatura têm verificado os impactos da entrada de uma empresa aérea de baixo custo (LCC) sobre a pontualidade dos voos no mercado, como Rupp (2008), Castillo-Manzano e Lopez-Valpuesta (2014), Bubalo e Gaggero (2015) e Prince e Simon (2015).

Nesse contexto, o presente trabalho tem como objetivo apresentar um estudo recente realizado no Núcleo de Economia do Transporte Aéreo do ITA que investigou os incentivos que a concorrência gera nas empresas aéreas dominantes para manterem a pontualidade dos voos na rota (nível local) e no aeroporto (nível global). A justificativa deste estudo é que a maioria dos artigos da literatura de atrasos aborda esse tema de forma isolada, focando apenas um desses níveis. Desta forma, o estudo apresentado neste trabalho desenvolveu uma estrutura unificadora para testar ambas as hipóteses de internalização do congestionamento no aeroporto e da qualidade do produto em um único modelo econométrico. O principal interesse do estudo é verificar os impactos que uma empresa aérea de baixo custo causa sobre as probabilidades e duração média dos atrasos das empresas aéreas incumbentes. Além disso, o estudo também investigou quais seriam os impactos da entrada local e global no que se refere à internalização do congestionamento e na qualidade do serviço prestado pela empresa aérea incumbente.

*Redes hub-and-spoke, dominância de aeroportos e atrasos*

O surgimento de redes hub-and-spoke é um fenômeno do período pós-desregulamentação do transporte aéreo dos Estados Unidos que se espalhou para a maioria dos mercados de empresas aéreas do mundo. A literatura científica tem observado que uma empresa aérea dominante em um aeroporto teria mais incentivos para lidar com o congestionamento do que as empresas aéreas menores e, portanto, naturalmente internalizaria os custos associados com seus próprios atrasos de voos (Daniel, 1995, Brueckner, 2002). Essa literatura foca no papel da alocação de voos e passageiros entre o pico/fora do pico no aeroporto para inspecionar os incentivos de gerenciar o congestionamento e evitar atrasos de voos das empresas aéreas dominantes. Nessa situação, a tarifa de congestionamento que visa mitigar as externalidades de atrasos de voos seria desnecessária ou apenas necessária em aeroportos menos dominados. Estudos recentes incluem Mayer e Sinai (2003) e Ater (2012).

A literatura vem investigando a hipótese da chamada "internalização do congestionamento no aeroporto". Essa hipótese diz que a concentração de mercado em um aeroporto geraria maiores incentivos para as empresas aéreas dominantes se envolverem na gestão dos atrasos, melhorando o desempenho de pontualidade – mas cobrando um preço por isso. "Internalizar a externalidade do congestionamento" seria, portanto, o fenômeno de uma empresa (ou poucas empresas) por conta própria cuidar do problema do congestionamento, aproveitando-se da sua dominância, para evitar danos à sua reputação e aos seus custos. Na vigência dessa hipótese, não seria necessário que a autoridade governamental ou regulador aplicasse algum tipo de taxação sobre voos ou demanda para aliviar o congestionamento.

Com relação à hipótese da internalização do congestionamento, a maioria dos trabalhos empíricos recentes encontrou uma relação negativa entre a concentração no aeroporto e atrasos de voos. Brueckner (2002) apresenta evidências rudimentares com base em uma amostra de 25 aeroportos nos Estados Unidos em 1999; os resultados são confirmados por Mayer e Sinai (2003) e Ater (2012), que usam dados desagregados em um painel de dados ao nível empresa aérea-rota-tempo para o mercado aéreo americano no início de 2000. Santos e Robin (2010) apresentaram uma aplicação para o mercado europeu de transporte aéreo entre 2000-2004 e confirmaram as conclusões de internalização do congestionamento. Por outro lado, Daniel e Harback (2008), Rupp (2009) e, em certa medida, Bilotkach e Lakew (2014), encontraram evidências de não-internalização, sugerindo que uma tarifa de congestionamento poderia melhorar a eficiência econômica. A métricas de concentração utilizadas nesses estudos são tipicamente o índice Herfindahl-Hirschman (HHI) no aeroporto ou a participação da empresa aérea dominante no aeroporto.

Outra vertente instigante da literatura de economia do transporte aéreo faz uma análise desagregada, ao nível de mercado – e não agregada ao nível do aeroporto – para inspecionar os determinantes dos atrasos de voos. O nível de mercado da indústria aérea está geralmente relacionado com o par origem-destino - o nível da rota. Enquanto a literatura de internalização do congestionamento no aeroporto está interessada em estimar os efeitos globais da concentração – ou seja, está preocupada com o nível do aeroporto – essa vertente da literatura está interessada em estimar os efeitos locais da concentração – ou seja, o nível de mercado (rota).

Ao nível da rota, alguns trabalhos recentes consideraram a pontualidade do serviço prestado como um dos principais indicadores da qualidade do serviço da empresa aérea. Os modelos empíricos de atrasos apresentam uma métrica de concorrência entre empresas aéreas no mercado como regressores além de outros fatores. A concentração de mercado é, nesse sentido, vista como um dos incentivos que as empresas aéreas têm para promoverem a qualidade do serviço. Suzuki (2000) apresenta pioneirismo nessa vertente ao mostrar que a pontualidade do serviço prestado afeta a participação de mercado na rota através da experiência dos passageiros relacionado aos atrasos. A literatura econométrica recente, no entanto, inverte a análise e estima uma equação de atrasos de voo contra a concentração de mercado (rota) - Mazzeo (2003), Rupp, Owens e Plumly (2006) e Greenfield (2014). Todos os artigos encontram evidências claras que suportam a relação positiva entre a qualidade do serviço prestado e concorrência do mercado. Temos assim, que, a partir dessa vertente da literatura, uma outra hipótese é levantada, a de relação entre qualidade e concorrência. Assim, a concentração na rota (mercado) geraria poucos incentivos para as empresas aéreas dominantes se envolverem na melhoria da qualidade do serviço prestado com relação a pontualidade dos voos.

Interessante de observar que a vertente da literatura que investiga a relação entre concorrência e qualidade do serviço não considera a possibilidade de internalização do congestionamento pelas empresas aéreas dominantes no aeroporto. De fato, nenhum dos artigos anteriores utilizam métricas de concentração ao nível do aeroporto para verificar os incentivos das empresas aéreas para fazer uma gestão mais eficaz dos atrasos – como por exemplo Mazzeo (2003), Rupp, Owens e Plumly (2006) e Greenfield (2014). Ao restringir a análise apenas ao nível de mercado, os estudos consideram que os atrasos de voos gerados pela concorrência são apenas gerados ao nível local da rota.

Uma outra questão associada à concorrência entre empresas aéreas e os incentivos para a gestão da pontualidade diz respeito às empresas aéreas de baixo custo (LCCs). Até o momento, poucos artigos da literatura concorrência-qualidade têm verificado os impactos da presença da LCC sobre os atrasos de voo. Rupp (2008) e Castillo-Manzano e Lopez-Valpuesta (2014) constataram que as LCCs apresentam uma melhor pontualidade no serviço prestado do que as companhias aéreas tradicionais. Entretanto, esse estudo apresenta uma abordagem semelhante à Bubalo e Gaggero (2015) e Prince e Simon (2015) que focam as respostas competitivas à entrada. Com base nessas pesquisas recentes, muitas perguntas podem ser formuladas para se entender melhor o funcionamento dos mercados aéreos:

- A entrada de uma novata LCC teria impactos sobre os incentivos das empresas estabelecidas em reduzirem os atrasos de voos?
- A entrada da LCC alteraria os incentivos das empresas estabelecidas em se engajar espontaneamente na internalização do congestionamento de um aeroporto em que dominam?
- A entrada dessa LCC ocasionaria uma maior concorrência pela qualidade do serviço prestado na rota, ou seja, uma maior pontualidade?

Até o momento, sobre essas questões, a literatura tem sido escassa e com resultados conflitantes. Prince e Simon (2015) encontraram evidências de que a entrada da LCC aumenta os atrasos de voos das empresas aéreas incumbentes, uma vez que ela forçaria uma concorrência de preços que induz tais empresas a cortar custos – um desses custos estaria associado à gestão da pontualidade do serviço prestado. No entanto, Bubalo e Gaggero (2015) encontram evidências contrárias. Portanto, o impacto da pressão concorrencial exercida pelas LCC sobre os atrasos de voos não apresenta consenso na literatura e ainda é uma questão empírica a ser investigada. Além disso, a literatura de internalização do congestionamento no aeroporto tem negligenciado a questão da entrada da LCC.

*Internalização do congestionamento*

A relação entre dominância no aeroporto por empresas aéreas e redução dos atrasos é conhecida na literatura como internalização do congestionamento. Ou seja, quando não há dominância no aeroporto, atrasos de voos não podem ser associados a nenhuma empresa aérea. Dessa forma, há tendência de que os passageiros percebam os atrasos mais como sendo um problema geral do aeroporto, ou do controle de tráfego aéreo, ou ainda um problema coletivo, e não como um problema de uma empresa aérea específica. Em teoria econômica, esse fenômeno é conhecido como "tragédia dos comuns": quando o aeroporto é muito compartilhado e não está fortemente ligada a alguma empresa aérea em particular, os problemas tendem a ser atrelados a ninguém e, por consequência, não há maiores incentivos para se cuidar deles ou mesmo evitá-los. Por outro lado, a chamada internalização do congestionamento ocorre quando uma ou poucas empresas dominam um aeroporto, isto é, possuem grande parcela dos voos. Em situações de dominância aeroportuária, a imagem das grandes operadoras tende a estar intimamente atrelada ao aeroporto e seus eventuais problemas. Atrasos nos voos do aeroporto acabam por manchar a reputação das operadoras dominantes e impactar sua função de custo demasiadamente, havendo assim uma tendência dessas empresas em empenharem maiores esforços em sua redução, alocando melhor seus voos entre horários de pico e fora do pico. Por isso, a gestão do risco de atrasos de voo tende, portanto, a ser mais eficaz em aeroportos naturalmente dominados.

A IATA mantém uma classificação por níveis de aeroportos. O Aeroporto Nível 1 (ou Non-Coordinated Airport) é aquele em que as capacidades de todos os sistemas do aeroporto estão adequadas para acomodar as demandas dos usuários. O aeroporto Nível 2 (ou Schedules Facilitated Airport) apresenta potencial para congestionamento em alguns períodos, mas é passível de resolução por meio de cooperação voluntária entre as companhias aéreas (sinal amarelo). Um schedules facilitator é designado para auxiliar nesse processo. Finalmente, o aeroporto Nível 3 (ou Coordinated Airport) apresenta elevado nível de congestionamento, tal que a demanda pela infraestrutura do aeroporto excede a capacidade nos períodos relevantes, mas as tentativas de solução de problemas operacionais (atrasos, cancelamentos) por cooperação voluntária, no Nível 2, falharam, sendo designado um Slot Coordinator independente.

### III. Caso dos Aeroportos Brasileiros

Para aplicação da teoria e das hipóteses acima discutidas, Bendinelli, Bettini e Oliveira (2016) investigam o problema com o objetivo de melhor entender o atrasos de voos na indústria aérea brasileira. Outro estudo que trata do tema, mas considerando o papel dos slots aeroportuários, é Miranda e Oliveira (2018)

No Brasil, atrasos de empresas aéreas têm sido discutidos há tempos, mas nenhuma tarifa de congestionamento foi implementada até o momento. Em 2008, três dos aeroportos mais importantes do país estavam entre os mais atrasados do mundo (Aeroporto Internacional de Brasília (BSB), Aeroporto de Congonhas (CGH) e o Aeroporto Internacional de Guarulhos (GRU)) . Para reverter essa situação, o país se envolveu em um grande esforço na supervisão de voos durante a Copa do Mundo de 2014, apresentando reformas na regulação para punir atrasos de voos longos e melhorar a gestão de slots escassos .

A aviação comercial no Brasil foi totalmente desregulamentada em 2001, com arranjos institucionais para a alocação de slots em aeroportos introduzidos pela primeira vez em 2006. Até agora, apenas o Aeroporto de Congonhas foi oficialmente designado como um aeroporto coordenado com rigorosas regras de alocação de slots ditadas pela Agência Nacional de Aviação Civil (ANAC), uma agência reguladora com autonomia e independência. O mecanismo de alocação é marcado pelo sistema de direitos adquiridos de "use-it-or-lose-it", mas com alguns regulamentos pró-novos entrantes . Os demais aeroportos não estão sujeitos as regras de alocação de slots. O Aeroporto Internacional de Guarulhos é atualmente um aeroporto com programação facilitada e participa da Worldwide Scheduling Guidelines and Conference da Associação Internacional de Transporte Aéreo (IATA).

Desde 2001, várias mudanças estruturais têm sido observadas na indústria aérea. Alguns exemplos são o surgimento das empresas aéreas de baixo custo como a Gol em 2001 e a Azul em (2008), a ascensão e a queda de uma importante aliança estratégica – o acordo de codeshare entre as empresas aéreas Varig e Tam em 2003-2005 -, e, mais recentemente, a privatização dos principais aeroportos desde 2012.

As grandes mudanças da aviação comercial através da década de 2000 produziram aspectos positivos e negativos. Entre 2002 e 2010, segundo a Agência Nacional de Aviação Civil o transporte aéreo doméstico apresentou um aumento de 153% na receita de passageiros por quilômetro voado e uma queda de 52% no valor médio pago por passageiro por quilômetro. Em paralelo, as empresas aéreas têm apresentado estratégias de marketing para atrair os consumidos, tais como a nova classe média emergente. Por outro lado, o rápido crescimento da demanda foi concomitante com a concentração de alguns principais hubs como os aeroportos da cidade de São Paulo, fazendo, assim, uma pressão sobre a infraestrutura aeroportuária existente. Após a falência da empresa aérea de bandeira Varig, a estrutura de mercado tornou-se rapidamente concentrada. A desconcentração do mercado recomeçou apenas em 2008, com a entrada da empresa aérea LCC Azul Linhas Aéreas, com aeronaves menores (E-Jets da Embraer) e uma intensa utilização de um nicho pouco explorado de operações no aeroporto secundário de São Paulo/Campinas (VCP).

O congestionamento nos aeroportos causados por gargalos de infraestrutura em todo o país foi evidente em toda a década de 2000 e, em particular, no "apagão aéreo" entre 2006-2007 -em que mais de um terço dos voos foram interrompidos- e, posteriormente, com a rápida aceleração do crescimento econômico a partir de 2010. A falta de concorrência nos aeroportos e o escasso orçamento público para melhorias e expansões mostraram que a escassez de infraestrutura aeroportuária ainda tem sido um problema no país. Desde o período do apagão, no entanto, atrasos de voos e cancelamento não tem sido um problema recorrente para as autoridades, uma vez que proporção de interrupções aéreas diminuiu consideravelmente. O estudo apresenta que a porcentagem de voos atrasados diminui consideravelmente ao compararmos o início da década de 2010 com qualquer período da década de 2000. Quando contrastado com a segunda metade da década anterior - que compreende o período de do apagão aéreo de 2006-2007 -, atrasos de voos diminuíram um terço (33,5%). Quando comparado com um período menos anormal, como os anos 2002-2005, a porcentagem de voos atrasados diminuiu em média 5,3%. Além disso, considerando a mesma comparação, o estudo aponta para uma redução de 11% no processo de "hubbing" - medido pela proporção de passageiros em conexão - e um aumento de 4,3% no HHI da cidade. Esta análise sugere que em alguns aeroportos a internalização do congestionamento pode ter ocorrido. Em contraste, o HHI do mercado (par de cidades) diminuiu 0,4%, sugerindo uma relação positiva leve entre concorrência e qualidade.

Bendinelli, Bettini e Oliveira (2016) analisam um conjunto de painel de dados de 209 rotas no Brasil entre 2002 e 2013. O conjunto de dados inclui apenas rotas envolvendo as capitais brasileiras e capital do país. Na análise, uma rota é definida como um par direcional de cidades domésticas. As empresas tradicionais da amostra correspondem à Tam, Varig, Transbrasil e Vasp, enquanto Gol e Azul são configuradas como LCCs. Adota-se a hipótese de que, pelo menos no período inicial de suas operações, tanto Gol quanto Azul apresentaram traços característicos de LCC – muito embora notáveis adaptações dos modelos de negócio tenham sido efetivadas ao longo do período. A Tabela 1 apresenta os resultados das estimações do modelo empírico de atrasos de voos desenvolvido pelo estudo. Note que reportamos apenas os sinais das estimativas obtidas pelos autores.

**Tabela 1 – Determinantes de atrasos de voo no Brasil**

| Variáveis | (1) Chance de Atrasos | (2) Minutos de Atraso | (3) Minutos de Atraso superior a 15 min |
|---|---|---|---|
| Número de voos em horários congestionados | + | + | + |
| Número de voos em horários não congestionados | + | NS | NS |
| Proporção de voos atrasados devido ao mau tempo | + | + | + |
| Proporção de voos atrasados devido a incidentes | + | + | + |
| Proporção de voos atrasados devido à conexão | + | + | + |
| Proporção máxima de voos atrasados | + | + | + |
| Acordos de codeshare | NS | NS | NS |
| Concentração de mercado (Índice HHI) na rota | + | + | + |
| Concentração de mercado (Índice HHI) no aeroporto | - | - | - |
| Presença de LCCs na rota | NS | + | + |
| Presença de LCCs no aeroporto | - | NS | NS |

Fonte: Bendinelli, Bettini e Oliveira (2016). "+", "-", e "NS" significam coeficiente estimado, respectivamente, positivo e estatisticamente significante, negativo e estatisticamente significante, e não estatisticamente significante.

Primeiro, em todos os modelos se têm uma evidência razoáveis da presença de internalização do congestionamento no aeroporto – a primeira hipótese do estudo, $H\_1$. De fato, os resultados para os coeficientes estimados de "HHI do aeroporto" são negativos e estatisticamente significantes em todas as especificações Chance de Atrasos e Minutos de Atraso. Tal fato confirma os resultados encontrados por Brueckner (2002), Mayer e Sinai (2003), Santos e Robin (2010) e Ater (2012). Segundo, em relação à hipótese qualidade-concorrência ($H\_2$), encontraram-se as evidências teóricas entre atrasos e concentração da rota de Mazzeo (2003) e Greenfield (2014). De fato, em todos os casos, os coeficientes do "HHI da rota" são positivos e estatisticamente significativos. Terceiro, sobre os impactos da entrada de uma empresa LCC, os resultados mostram alguma evidência de que as empresas tradicionais envolveram-se em uma internalização adicional de atrasos após a entrada no que se refere à probabilidade de atrasos nos voos, mas não para a duração desses atrasos, uma vez que os coeficientes negativos de "Presença de LCCs no aeroporto" são estatisticamente significativos na equação Chance de Atrasos, mas não são na equação Minutos de Atraso. As respostas a entrada ao nível local (rota), examinadas com a presença da empresa LCC, não são significativas na especificação Chance de Atrasos. Portanto, não foram encontradas evidências que suportem a hipótese de redução de custos/preços de Prince e Simon (2015).

Os resultados sugerem a conclusão de que a internalização do congestionamento nos aeroportos pode ser observada no mercado brasileiro de aviação e também é induzida pela entrada da LCC. Após a entrada, a participação na cidade da empresa aérea dominante diminui no modelo Minutos de Atraso e a consequente queda na concentração tende a provocar uma redução na internalização do congestionamento e um aumento nos atrasos de voos. Nesse sentido, as estimativas mostram que a presença da LCC na cidade apresenta efeito moderador sobre a redução da internalização do congestionamento. Uma estratégia de "depeaking" (descongestionamento de horários de pico em prol de horários fora de pico) pode ser utilizada como uma tentativa de manter a internalização do congestionamento mesmo com a diminuição da concentração do aeroporto através da redução da complexidade das operações hub-and-spoke ao competir com a LCC, por exemplo. Dessa forma, voos podem ser alocados para horários fora do pico onde a LCC é mais atraente para os passageiros a lazer – fins de semana, por exemplo. Outra estratégia alternativa que visa melhorar a pontualidade do serviço também é possível para justificar os resultados. Por exemplo, melhorias não observadas na rede e no gerenciamento da programação das empresas incumbentes após a entrada da LCC pode induzir reduções permanentes de atrasos de voos. Tais resultados são consistentes com o movimento global de suavização do pico de operações da companhia tradicional para reduzir o congestionamento global. A entrada em uma rota, portanto, gera concorrência potencial sobre as outras rotas de uma cidade e, portanto, resulta em um "efeito repercussão" (spillover) que beneficia rotas em que ela não entrou em direção à melhoria da pontualidade. Note, no entanto, que não foram encontradas evidências suficientes para apoiar qualquer efeito moderador ou um efeito intensificador da entrada da LCC na relação qualidade-concorrência. Dessa forma, não foi observado qualquer efeito ceteris paribus da entrada da LCC sobre os atrasos de voos ao nível da rota, além do efeito global da internalização adicional provocada pela entrada sobre a prevalência dos atrasos.

## IV. Conclusões

O presente trabalho visou estudar a questão de atrasos de voo na aviação comercial. Resenhamos um estudo nacional que estimou tanto os efeitos locais quanto os globais da concorrência na pontualidade das empresas aéreas estabelecidas tradicionais no mercado brasileiro de aviação. Através do desenvolvimento de modelos econométricos de probabilidades e de duração dos atrasos de voos, Bendinelli, Bettini e Oliveira (2016) testaram relações importantes encontradas na literatura atual como as hipóteses de internalização do congestionamento no aeroporto e a associação entre a qualidade de serviço prestado e concorrência no mercado. Avaliaram-se também os impactos da concorrência efetiva e potencial de uma empresa aérea de baixo custo (LCC) na indústria.

Os resultados dos autores sugerem que a internalização do congestionamento no aeroporto foi observada no mercado aéreo brasileiro no período analisado. Além disso, também foi induzida pela entrada LCC sendo que o efeito da concorrência potencial causa uma repercussão positiva que beneficia rotas não entradas através da melhoria da pontualidade. No entanto, há evidencias que a entrada da LCC beneficia os consumidores pela redução da prevalência dos atrasos de voos, mas sem evidências sobre o impacto na duração de tais atrasos. Encontraram-se evidências suficientes para apoiar a hipótese de qualidade-concorrência em que uma menor concentração ao nível da rota melhoraria a qualidade do serviço prestado pela empresa aérea e forçaria, assim, os atrasos de voos a declinarem.

A interpretação dos resultados combinados dos efeitos locais e globais das condições de concorrência no comportamento das empresas aéreas incumbentes em relação à pontualidade, encontrados no estudo dos autores, pode ser resumida da seguinte maneira. Em um aparente paradoxo, as empresas tendem a internalizar o congestionamento quando seu domínio no aeroporto é aumentado, mas também tendem a manter alguma internalização quando esse domínio é desafiado pela entrada de uma empresa aérea do tipo LCC. Estes movimentos não são contraditórios, pois podem ser resultado de ajustes de programação estratégicos provocados pela antecipação da forte concorrência de preços com um recém-chegado que apresenta um modelo de negócio diferente. Portanto, os efeitos locais e globais combinados indicam que, embora a concorrência por qualidade no que se refere à pontualidade é observada localmente no mercado, há evidências de que o surgimento e crescimento de LCCs podem ser um fator adicional para a melhoria na pontualidade da indústria aérea. No entanto, há certamente uma necessidade de considerar as aplicações desse modelo em outras regiões e também considerar diferentes categorias de LCCs em todo o mundo.